\documentclass[showpacs,aps,prb,twocolumn,groupedaddress]{revtex4}

\usepackage{graphicx}

\begin{document}


\title{Low temperature vortex liquid states induced by quantum fluctuations in the quasi two dimensional organic superconductor $\kappa$-(BEDT-TTF)$_{2}$Cu(NCS)$_{2}$}
\author{T. Sasaki}
\author{T. Fukuda}
\author{T. Nishizaki}
\author{T. Fujita}
\author{N. Yoneyama}
\author{N. Kobayashi}
\affiliation{Institute for Materials Research, Tohoku University, Katahira 2-1-1, Sendai 980-8577, Japan}
\author{W. Biberacher}
\affiliation{Walther-Meissner-Institute, Walther-Meissner Str. 8, D-85748 Garching, Germany}
\date{\today}

\begin{abstract}
We report the transport properties in the vortex liquid states induced by quantum fluctuations at low temperature in the layered organic superconductor $\kappa$-(BEDT-TTF)$_{2}$Cu(NCS)$_{2}$.  
A steep drop of the resistivity observed below about 1 K separates the liquid state into two regions.  
In the low resistance state at lower temperature, a finite resistivity with weak temperature dependence persists down to 100 mK at least. 
The finite resistivity in the vortex state at $T \simeq$ 0 K indicates the realization of quantum vortex liquid assisted by the strong quantum fluctuations instead of the thermal one.  
A possible origin for separating these liquid states is a remnant vortex melting line at the original position, which is obscured and suppressed by the quantum fluctuations. 
A non-linear behavior of the in-plane resistivity appears at large current density in only the low resistance state, but not in another vortex liquid state at higher temperature, where the thermal fluctuations are dominant.  
The transport properties in the low resistance state are well understood in the vortex slush concept with a short-range order of vortices.  
Thus the low resistance state below 1 K is considered to be a novel quantum vortex slush state.
\end{abstract}

\pacs{74.70.Kn, 74.40.+k, 74.60.-w}

\maketitle

\section{Introduction}

Vortices in the layered superconductors (i.e., high-$T_{\rm c}$ superconducting oxides, organic superconductors) have strong thermal fluctuations, which have been extensively studied. \cite{Blatter94A,Nishizaki00}  
The material parameters of such layered superconductors, implying their short coherence length and large anisotropy of the effective mass, enhance the importance of the fluctuation effects on many supercondicting phenomena such as the thermodynamic and transport properties, and also on the vortex matter dynamics and its phase diagram.
Quantum fluctuations on the superconductivity and the vortices are expected to become potentially important at low temperatures on the materials fairly affected by the thermal fluctuations. \cite{Ikeda96}  
Indeed the vortex liquid state resulting from the quantum melting at low temperature has been discussed as the quantum vortex liquid (QVL) from several theoretical approaches. \cite{Ikeda96,Blatter93,Blatter94B,Chudnovsky,Rozhkov,Onogi,Kramer} 
 The favorable material parameters for the experimental observation of quantum fluctuation effects involve a large normal-state resistivity $\rho_{\rm n}$, a moderate upper critical field $H_{\rm c2}$ at zero temperature, and a small length scale $s$ for the fluctuations (i.e., short coherence length or short layer separation).  

Besides oxides, the BEDT-TTF molecule based organic superconductors are also good candidates for the observation of these effects, where BEDT-TTF denotes bis(ethylenedithio)tetrathiafulvalene.  
The quasi two dimensional (Q2D) organic superconductors have relatively large effective normal resistance $R_{\rm eff} = \rho_{\rm n}/s$ in comparison with the quantum resistance $R_{\rm Q} = \hbar/e^{2}$.  
For example, $\kappa$-(BEDT-TTF)$_{2}$Cu(NCS)$_{2}$, which is investigated in the present study, shows the resistance ratio $Q \equiv R_{\rm eff}/R_{\rm Q}$ of the order of 10$^{-1}$, rendering the quantum effects important. \cite{Blatter94A,Blatter94B} 
Further the fairly large Ginzburg number $Gi \equiv (1/2)(T_{\rm c}/H_{\rm c2}^{2}(0){\varepsilon}{\xi}^{3}(0))^{2}$ of the order of 10$^{-1}$ in $\kappa$-(BEDT-TTF)$_{2}$Cu(NCS)$_{2}$ is in favor of the thermal and also the quantum fluctuations, \cite{Blatter94A,Ikeda96} whereas $Gi$ in conventional superconductors is a small number of the order of 10$^{-7}$, and in oxides, YBa$_{2}$Cu$_{3}$O$_{y}$ (YBCO), of the order of 10$^{-2}$. 
In the case of the strongly layered oxides, Bi$_{2}$Sr$_{2}$CaCu$_{2}$O$_{y}$ (BSCCO) possesses $Gi \sim$ 10$^{0}$.  

In addition to the material potential for the quantum fluctuations, it is practically important that a moderate $H_{c2}(0) \simeq$ 6 $-$ 7 T in the magnetic field perpendicular to the Q2D plane can be accessed fully by using standard superconducting magnets.  
Indeed the QVL state in $\kappa$-(BEDT-TTF)$_{2}$Cu(NCS)$_{2}$ has been found  as a reversible magnetization region below $H_{\rm c2}$ even at $T \simeq$ 0 K. \cite{Sasaki98,Mola01}

The transport properties in the QVL state are expected to show some characteristic phenomena such as an insulating behavior. \cite{Ikeda96,Ikeda96B}  
The experimental efforts, however, have not been enough to reveal the properties in the QVL state.  
Most of works have been performed on thin films of conventional superconductors because of meeting the requirement mentioned above. \cite{Ephron,Kes,Markovic,Chervenak,Okuma}
The thin films, however, are not so clean that the wide vortex liquid region is not anticipated by the large random pinning force on the vortices.  
On the other hand, the organic superconductor obtained as a single crystal is so clean that the vortex phase diagram has been discussed on the resemblance with that of BSCCO, \cite{Nishizaki96,Lee} whereas their $T_{\rm c}$'s are different by one order of magnitude.  
Furthermore the cleanness of the materials enables us to obtain the electronic states in detail by measurements of the magnetic quantum oscillations, de Haas - van Alphen (dHvA) and Shubnikov - de Haas (SdH) effects in the normal \cite{Oshima,Sasaki90,Harrison} and also the superconducting \cite{Sasaki98,Wel,Clayton,Sasaki02A} states.  

In this paper, we report the transport properties of the title organic superconductor at low temperature, where the QVL state is expected.  
We are not concerned here with the debatable nature of the superconductivity reported in the $\kappa$-type of BEDT-TTF based organic superconductors, while those are currently discussed on the pairing symmetry and the gap structure \cite{Izawa,Mueller02A} in connection with the electronic phase diagram in the normal state. \cite{Kanoda,Sasaki02B,Mueller02B} 

\section{Experiment}

High quality single crystals of $\kappa$-(BEDT-TTF)$_{2}$Cu(NCS)$_{2}$ were grown by an electrochemical oxidation method. 
The crystals measured have the shape of an elongated hexagonal plate with the typical size of $\sim$ 1.5 $\times$ 0.6 $\times$ 0.05 mm$^{3}$. 
The in-plane and the out of plane resistivities were measured along the $b$ and $a^{*}$ axes, respectively, by means of a conventional ac or dc four terminal method.  
The electrical terminals were made of evaporated gold films, and gold wires (10 ${\mu}$m) were glued onto the films with gold or silver paint.  
The contact resistance was about 10 $\Omega$ for each contact at room temperature, but it became less than 1 $\Omega$ at low temperature where the experiments were carried out.
The results presented in this paper were obtained on three samples \#1, \#2 and \#3 from different batches.  
We found that other three samples measured gave qualitatively similar results which were not presented in this paper.   
These samples were cooled slowly from room temperature to 4.2 K in 24 $-$ 48 hours in order to avoid the disorder of the terminal ethylene group of the BEDT-TTF molecules.  
The samples were directly immersed in liquid $^{3}$He or the dense $^{3}$He gas of the refrigerator within a 15 T superconducting magnet at the High Magnetic Field Laboratory for Superconducting Materials (HFLSM), IMR, Tohoku University. 
The sample \#3 was measured by using the $^{3}$He/$^{4}$He dilution refrigerator with a 14 T superconducting magnet at Walther-Meissner-Institute (WMI).  
For all measurements reported here, the magnetic field was applied parallel to the $a^{*}$ axis, i.e., the perpendicular to the Q2D conducting $b$-$c$ plane.  

\section{Results and discussion}

\begin{figure}
\includegraphics[viewport=3cm 3cm 19cm 27cm,clip,width=0.9\linewidth]{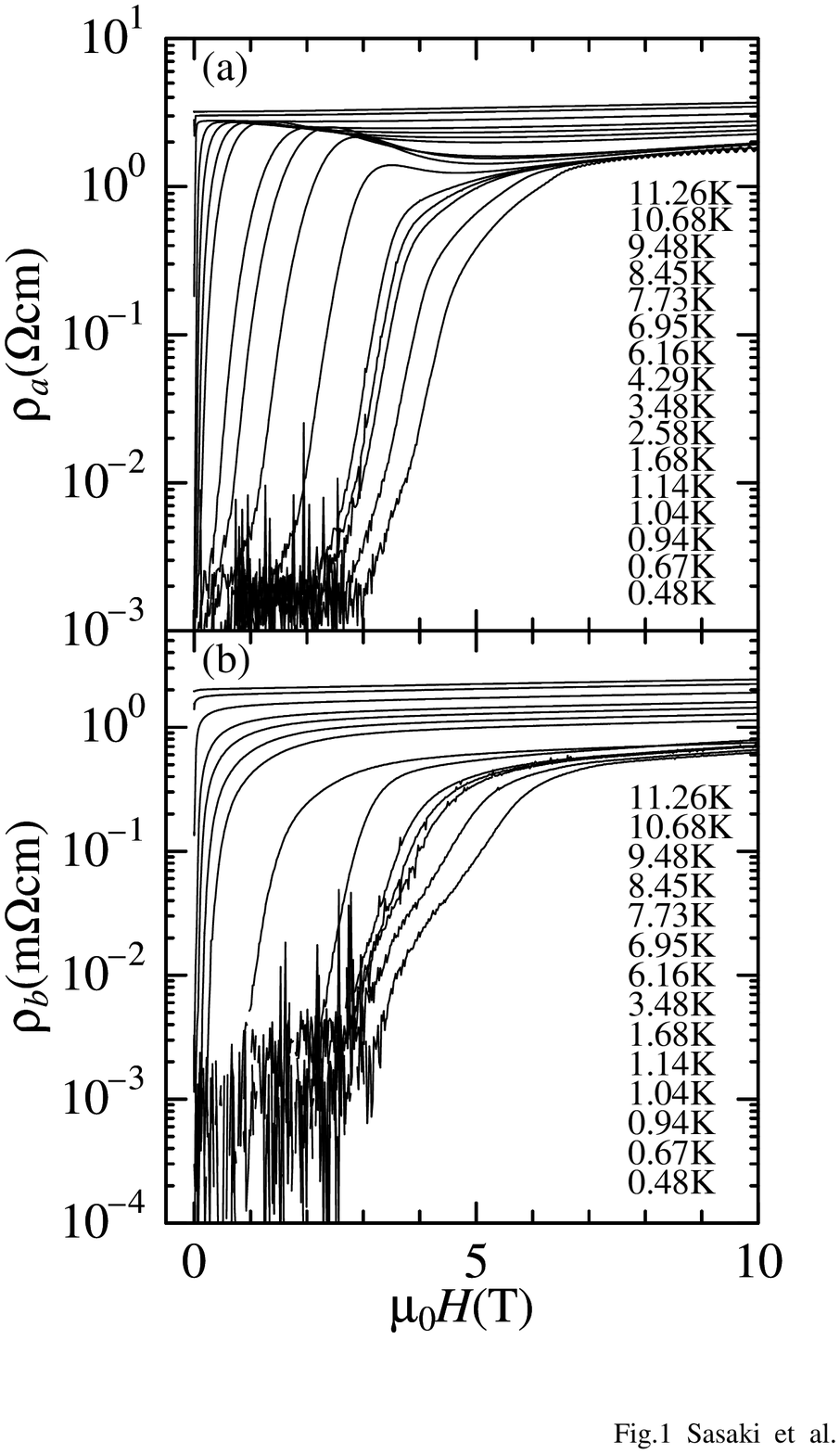}
\caption{Magnetic field dependence of the resistivity of $\kappa$-(BEDT-TTF)$_{2}$Cu(NCS)$_{2}$ in the magnetic field perpendicular to the Q2D plane.  (a) The out of plane resistivity $\rho_{a}$ along the $a^{*}$ axis ($I = 10 {\mu}$A), and (b) the in-plane resistivity $\rho_{b}$ along the $b$ axis ($I = 100 {\mu}$A) are measured in sample \#1.}
\end{figure}

Figure 1 shows the magnetic field dependence of the resistivity (a) $\rho_{a}$ along $a^{*}$ axis and (b) $\rho_{b}$ along $b$ axis of sample \#1 with dc current $I$ of 10 and 100 $\mu$ A, respectively.  
An anomalous resistance hump is observed in $\rho_{a}(H)$ curves at temperatures between about 1.5 K and $T_{c} \simeq$ 10 K.  
The resistivity hump is larger than the isothermal normal resistivity in high magnetic field.  
In the corresponding temperature - magnetic field region, the in-plane resistivity $\rho_{b}$ does not show such anomaly. 
This hump anomaly has been reported on only the interlayer resistivity of this material and discussed in several ways. \cite{Pratt,Ito,Friemel,Zuo,Kartsovnik99} 
But a persuasive explanation has not been presented.  
Although the purpose of this paper is not on the resistivity hump, it may be a clue for understanding the problem that the phenomena occur in the thermal vortex liquid (TVL) state which is described latter.  

The $\rho_{b}$ curve shows gradual decrease from the normal resistivity above $H_{\rm c2}$ with decreasing magnetic field.  
This gradual decrease of the resistivity demonstrates the dissipations due to the flux motion, resulting in the vortex liquid state.  
Further decreasing field, the resistivity becomes zero at which the long-range order of vortices starts to grow and the resulting vortex lattice (the vortex solid) is pinned by the random pinning potential like defects and impurities.  
Although the solid - liquid transition has been suggested to be a first-order vortex melting one at low field region ($\sim$ several ten mT), \cite{Inada} magnetic measurements in higher fields have obtained those only as the irreversible field $H_{\rm irr}$. \cite{Sasaki98,Nishizaki96} 
And the transition at the resistance onset is not so sharp as against the melting transition. 
It is not clear whether the zero resistivity at the field of tesla order comes in by the first-order melting or the second-order glass - liquid transition.  

A step-like structure is found in the $\rho_{b}$ curves on the way of the gradual transition in the vortex liquid region below 1 K. 
It means that a low resistance state appears before achieving the zero resistivity.
The features of the resistivity and the resulting low resistance state is found to involve an applied current density dependence.

\begin{figure}
\includegraphics[viewport=2cm 7cm 20cm 23cm,clip,width=0.9\linewidth]{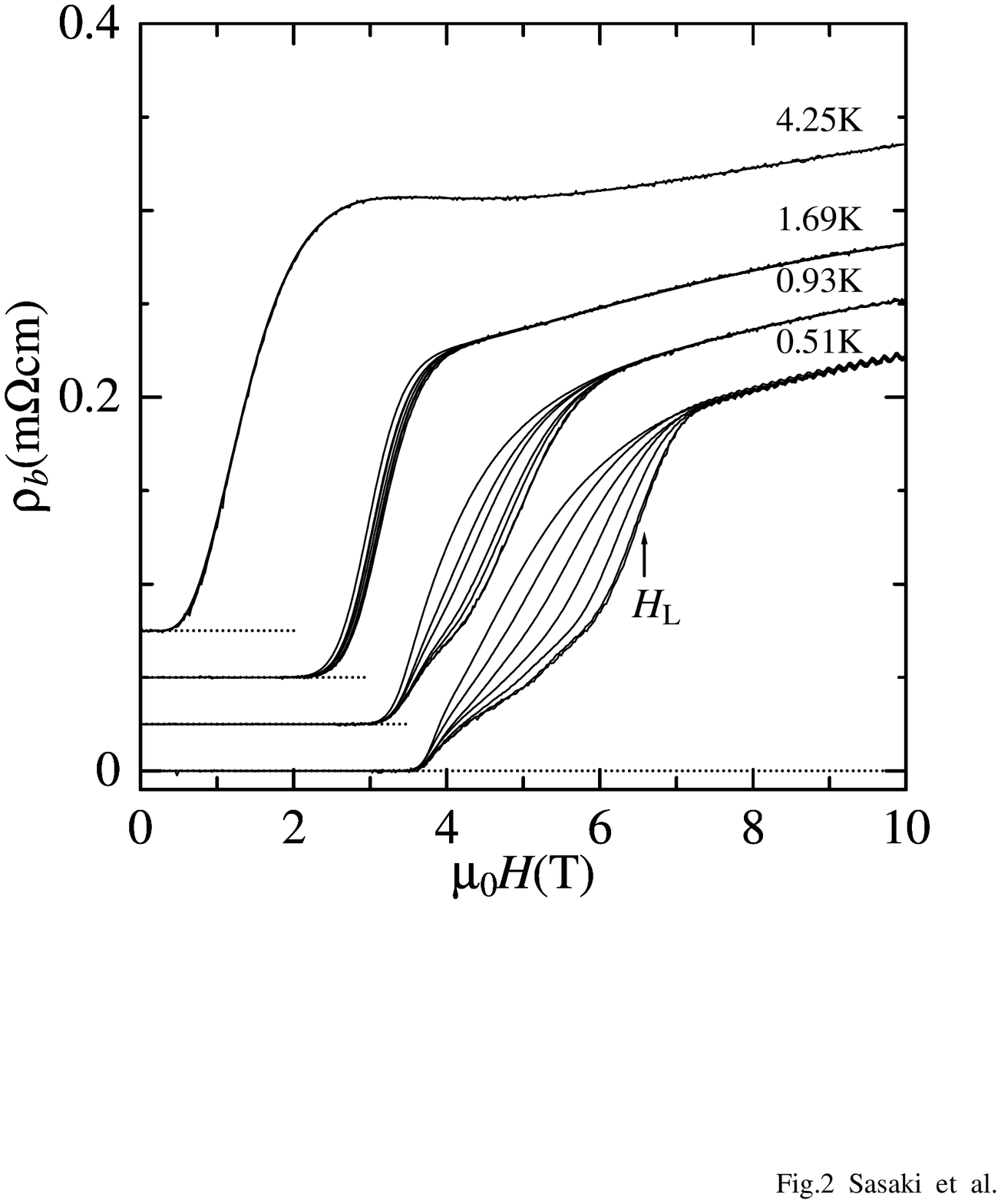}
\caption{Magnetic field dependence of the in-plane resistivity $\rho_{b}$ of sample \#2 with several current densities in the filed perpendicular to the Q2D plane.  The applied dc currents are 500, 200, 100, 50, 20, 10 and 5 $\mu$A from top to bottom curves at 0.51, 0.93 and 1.69 K, and 500, 100 and 10 $\mu$A at 4.25 K.  Current density $J$ corresponds to the applied current in the ratio of 1.53 A/cm$^{2}$ to 500 $\mu$A.}
\end{figure}

Figure 2 shows the magnetic field dependence of the in-plane resistivity $\rho_{b}$ of the sample \#2 in the field perpendicular to the plane with different dc current density.  
The applied currents $I$ at 0.51, 0.93 and 1.69 K are 500, 200, 100, 50, 20, 10 and 5 $\mu$A, and 500, 100 and 10 $\mu$A at 4.25 K from top to bottom curves at each temperature. 
The current density $J$ corresponding to 500 $\mu$A is 1.53 A/cm$^{2}$.  
The transition curves below about 1 K indicate large non-linear resistance behavior.  
In low current density, for example at 0.51 K, a steep resistance drop appears at $H_{\rm L} \simeq$ 6.5 T.  
The low resistance state following the resistance drop continues down to the field where the resistivity becomes zero.  
With increasing current density, however, the feature of the resistance drop becomes unclear and the region of the low resistance state shrinks.  
No such resistance drop at $H_{\rm L}$, and consequently no low resistance state are recognized at 1.69 K.  
It is noted that the observed current density dependence is not caused by the generation of heat at the sample and electrical contacts.  
Because the resistive onset field $H_{\rm irr}$ is not affected by changing current density, and also the resistive transition curves at 4.25 K, which are measured with the same current density range, coincide with each other.  

\begin{figure}
\includegraphics[viewport=2cm 8cm 20cm 24cm,clip,width=0.9\linewidth]{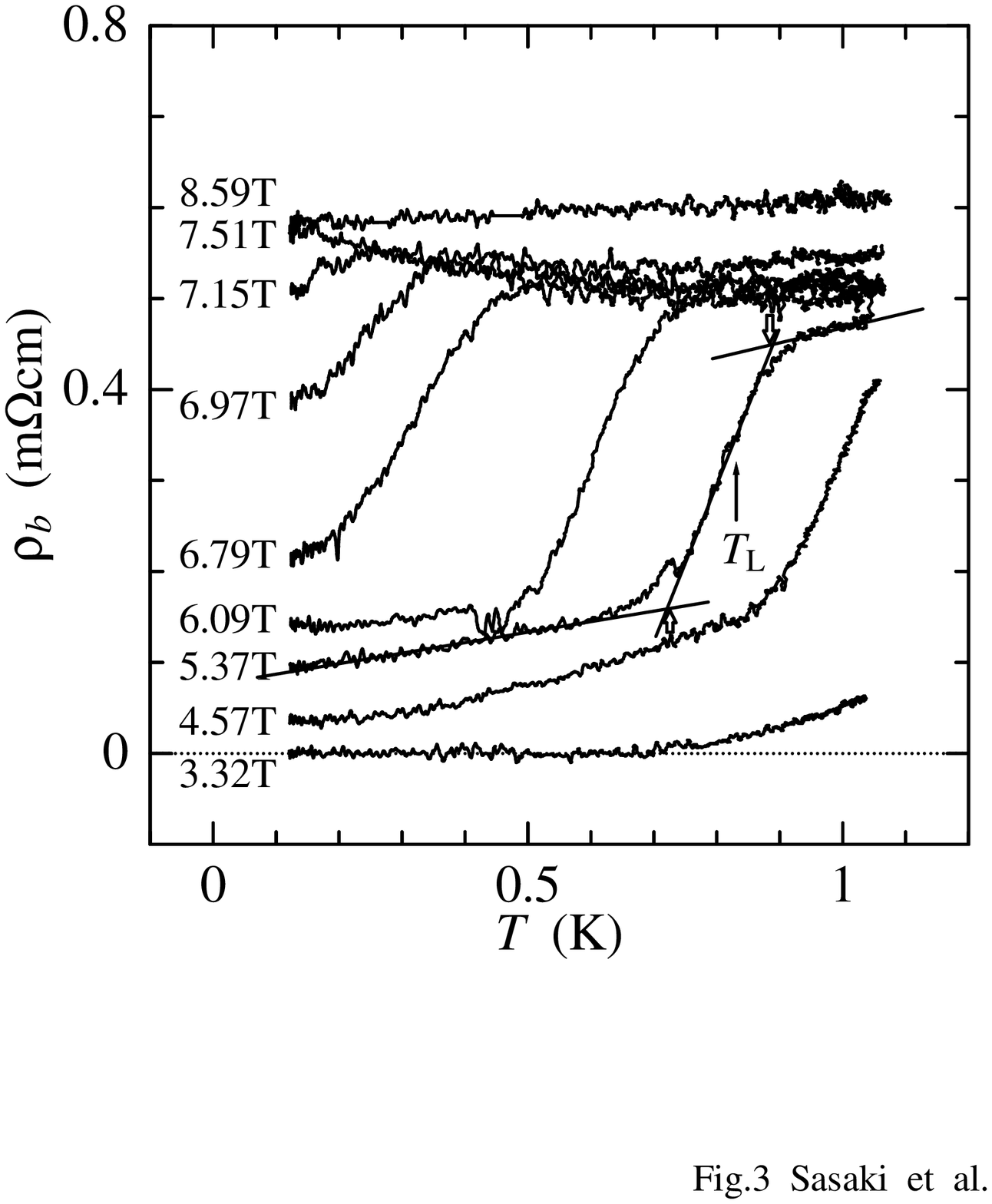}
\caption{Temperature dependence of the in plane resistivity $\rho_{b}$ in the field perpendicular to the Q2D plane of sample \#3. The arrows indicate a transition at $T_{\rm L}$ and the width between the high and low resistance states in the vortex liquid state, which are plotted as the bar in the phase diagram of Fig. 4.}
\end{figure}

The resistance drop at $H_{\rm L}$ and following low resistance state are more clearly seen in the temperature dependence of the resistivity at temperatures below 1 K.  
Figure 3 shows the temperature dependence of the in-plane resistivity $\rho_{b}$ of the sample \#3.  
The resistivity measurements were performed by an ac current (77 Hz) of 20 $\mu$A ($j =$ 0.05 A/cm$^{2}$).  
The current density used in the measurement is almost the same level of that in the two resistivity curves (10 and 20 $\mu$A) measured on the sample \#2 in Fig.2.  
The resistance drop at $T_{\rm L}$ corresponding to $H_{\rm L}$ has relatively narrow transition width of about 200 mK, which is defined by the intersection points of the linear extrapolation lines to the resistance curve as depicted in Fig. 3. 
On the other hand, the zero resistivity transition seems to be gradual as is seen on the curve in 3.32 T and at $T =$ 0.6 $-$ 0.7 K.  
The gradual transition to the zero resistivity may show a second-order vortex liquid - glass transition although the transition curve can not be analyzed by the glass scaling law \cite {Fisher89,Fisher91} because of insufficient accuracy of the resistance measurements near zero resistivity.  
The transition curve at $T_{\rm L}$ shifts almost  in parallel to lower temperature with increasing magnetic field, and the feature of the resistance drop can be seen up to nearby $H_{\rm c2}(0) \simeq$ 7 T.  
The low resistance state following the resistance drop persists down to about 100 mK at least.  
In addition the temperature dependence of $\rho_{b}$ in the low resistance state is relatively weak.  
Then we expect that the low resistance state could remain even at $T \simeq$ 0 K between $\sim$ 3.5 and $\sim$ 7 T.  
This finite resistivity is in agreement with the reversible magnetization \cite{Sasaki98} observed in the QVL state affected strongly by the quantum fluctuations.  
The quantum melting transition reflected on $H_{\rm irr}$ between the vortex solid and the QVL states at $T \simeq$ 0 K has been discussed quantitatively \cite{Sasaki98} in comparison with the several quantum melting theories. \cite{Ikeda96,Blatter93,Blatter94B,Chudnovsky,Rozhkov}  
Using the material parameters of the present organic superconductor, the quantum melting transition fields have been calculated to be 3.5 - 4.0 T at $T =$ 0 K by different theoretical ways.

It is expected that the quantum fluctuation effect may be seen in the resistance behavior. \cite{Ikeda96,Onogi}  
At lower temperature, $\rho_{b}$ above $T_{\rm L}$ shows a weak upward curvature, whereas the curve at 8.59 T, where the magnetic field is well above $H_{\rm c2}$, does not show such feature.
This may indicate an insulating behavior predicted as strong quantum fluctuation effect at $T \simeq$ 0 K and near $H_{\rm c2}$. \cite{Ikeda96,Ikeda96B} 
Theoretically the resistivity at $T =$ 0 K is expected to take either zero or the normal state value. \cite{Ikeda96} 
It means that either the insulating behavior results in taking the normal state value, or the finite pinning in the real system does in achieving zero resistivity.  
The intermediate metallic behavior with weak temperature dependence of the resistivity curves from 4.57 T to 7.15 T shows the tendency of neither insulating nor taking zero value down to 0.1 K.
This is similar to the observation of the metallic quantum vortex liquid state found in thick amorphous Mo$_{x}$Si$_{1-x}$ films at $T \simeq$ 0 K. \cite{Okuma}
In order to confirm this point in the present organic superconductor, further experiments at lower temperatures and in finely tuned magnetic fields are required.  

\begin{figure}
\includegraphics[viewport=2cm 7.5cm 20cm 25cm,clip,width=0.9\linewidth]{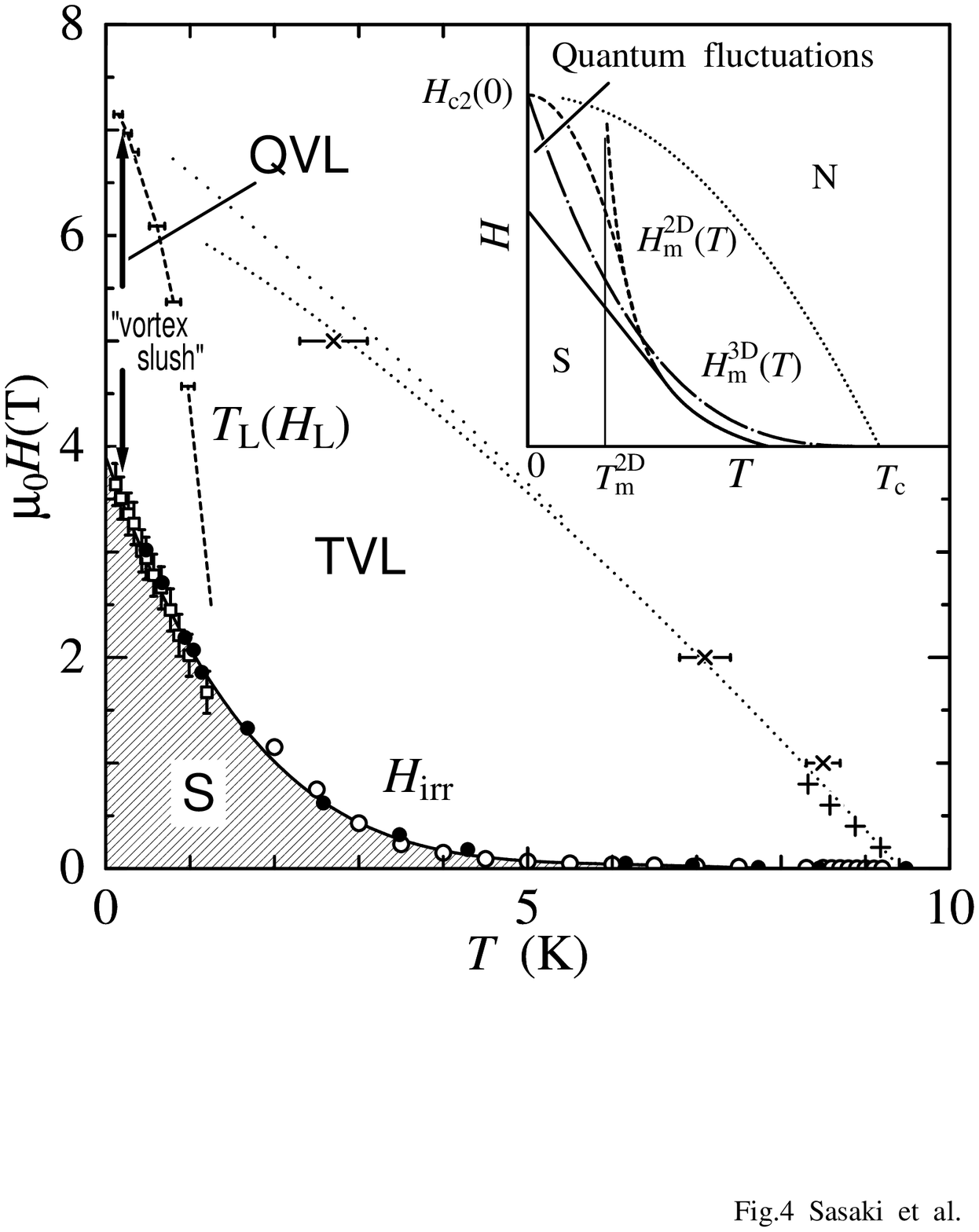}
\caption{Vortex phase diagram of $\kappa$-(BEDT-TTF)$_{2}$Cu(NCS)$_{2}$ in the magnetic field perpendicular to the Q2D plane.  QVL and TVL are the quantum and thermal vortex liquid state. N and S denote the normal and the vortex solid states.  The lower dotted curve shows the ordinary $H_{\rm c2}$ including the effect of fluctuations.  The curve is determined by the specific heat \cite{Graebner} (pluses) and the magnetization \cite{Lang94} (crosses) measurenents. The expected mean field $H_{\rm c2}$ line is plotted  by another dotted curve (upper side).  $H_{\rm irr}$ indicates the irreversible magnetic field obtained by the magnetization \cite{Nishizaki96} (open circles), the magnetic torque \cite{Sasaki98} (open squares) and the resistivity measurements in the present study (filled circles).  The dashed line $T_{\rm L}$ ($H_{\rm L}$) separates the vortex liquid state to two regions; the thermal vortex liquid (TVL) state with high resistivity at high temperature and the vortex slush like low resistance state at low temperature.  The inset shows a schematic phase diagram of the layered superconductors. The detail see the text.}
\end{figure}

The experimentally obtained $H_{\rm L}$, $T_{\rm L}$ and the low resistance state with non-linear behavior apparently show that the vortex liquid state of this material is separated into two kinds of the vortex liquid; one is the thermally fluctuated vortex liquid (TVL) at high temperature and another the non-linear low resistance vortex slush \cite{Worthington} like state where the quantum vortex liquid (QVL) may be realized by the strong quantum fluctuation instead of the thermal one at low temperature.

Figure 4 shows the phase diagram of vortex system in the magnetic field perpendicular to the Q2D plane of $\kappa$-(BEDT-TTF)$_{2}$Cu(NCS)$_{2}$.  
In the main panel the hatched low-$T$, low-$H$ region is the vortex solid state (S), which is characterized by the zero resistance and the irreversible magnetization. 
The squares and circles are the irreversible field $H_{\rm irr}$ determined by the magnetic torque \cite{Sasaki98} and SQUID \cite{Nishizaki96} measurements, respectively.  
The filled circles are determined as the onset magnetic field of $\rho_{a}$ in Fig. 1(a).  
The resistivity level of 2 $\times$ 10$^{-3}$ ${\Omega}$cm is used for the criterion of the onset.  
The resistive onset magnetic fields for $\rho_{b}$ agree with those for $\rho_{a}$ by extrapolating the $\rho_{b}$ data to the same criterion level (6 $\times$ 10$^{-4}$ m$\Omega$cm) which is 10$^{-3}$ times smaller than the normal state value.  
The pluses and crosses are the ordinary upper critical field $H_{\rm c2}$ obtained by the specific heat \cite{Graebner} and the magnetization \cite{Lang94}.  
The lower dotted curve of the ordinary $H_{\rm c2}$ on the pluses and crosses is expected to be lowered from the mean field $H_{\rm c2}$ (the upper dotted curve) due to fluctuations. \cite{Ikeda96}
Above the $H_{\rm irr}$ line in the TVL region, the vortices do not have any long-range order resulting in the vortex liquid state.  
Most of the liquid state (TVL) above several hundred mT is considered to be formed by melting or decoupling the vortex pancake \cite{Nishizaki96,Lee} in the S state, which is mainly assisted by thermal fluctuations.  
In the very low magnetic field region below the dimensional crossover at several ten mT, the Abrikosov vortex line lattice exists. \cite{Lee,Vinnikov}
On the low magnetic field phases which are not included in Fig. 4, please refer to the previous reports. \cite{Nishizaki96,Lee}

The broken line ($T_{\rm L}$ and $H_{\rm L}$) separates the vortex liquid state into two regions mentioned above.  
One is the TVL state at higher $T$, and another the non-linear low resistance state at lower $T$.  
This low resistance state persists down to at least 100 mK in $H_{\rm irr} < H < H_{\rm c2}$ at $T \simeq$ 0 K.  
The transport properties in the low resistance state below $H_{\rm L}$ resemble those observed in the vortex slush phase which has been found between the vortex liquid and glass phases of the high-$T_{\rm c}$ oxide superconductors having an intermediate range of disorder. \cite{Worthington,Nishizaki00A}  
The vortex slush phase has the short-range order of the vortices, \cite{Worthington,Ikeda01,Nonomura} which is characterized by the non-linear resistivity. \cite{Worthington,Nishizaki00A,Wen,Shibata02} 
The transition between the vortex liquid and slush phases appears as a steep resistance drop but not to zero, and also a small magnetization jump. \cite{Nishizaki00A,Shibata02} 
This transition is thought to be an incomplete first-order phase transition at the same transition field and temperature of the original one hidden by the disorder effect.  
In addition the second-order vortex glass transition line appears between the vortex slush and glass phases.  

Let us discuss the correspondence between the non-linear low resistance state found in this material and the vortex slush phase with a short-range order of vortices. 
The inset of Fig.4 shows a schematic phase diagram of the layered superconductors. \cite{Blatter94A,Tinkham} 
In the case of the 3D material, e.g., YBCO, the melting line (the dotted chain) points to $H_{\rm c2}$(0) because of less effect of fluctuations.  
In contrast to the 3D system, the melting line (the broken line) of the 2D material, e.g., BSCCO and the present organic superconductor, increases rapidly near the 2D melting temperature $T_{m}^{\rm 2D}$, which is independent of magnetic field at the Berezinskii-Kosterlitz-Thouless (BKT) type dislocation-mediated melting transition \cite{Huberman}, $kT_{m}^{\rm 2D} = d{\varepsilon}_{0}/8\sqrt{3}\pi$, where ${\varepsilon}_{0} = ({\Phi}_{0}/4\pi{\lambda})^{2}$, $k$ is the Boltzmann's constant, $d$ the layer spacing, $\Phi_{0}$ the flux quantum and $\lambda$ the in-plane penetration depth.  
The magnetic field independent $T_{m}^{\rm 2D}$ is expected at 1.5 - 2.5 K in the present organic superconductor with $d \simeq$ 15.2 {\AA} \cite{Watanabe} and the in-plane $\lambda \simeq$ 5000 - 7000 {\AA}. \cite{Lee,Vinnikov,Lang94,Lang92} 
In the real material the 2D melting line is expected to get away from $T_{m}^{\rm 2D}$ and turn toward $H_{\rm c2}$(0) in high magnetic field.

The $T_{\rm L}$ ($H_{\rm L}$) line observed in this material seems to coincide with the above mentioned melting line $T_{\rm m}^{\rm 2D}$, which corresponds to the dislocation mediated melting line \cite{Blatter94B}, following the broken line in the inset.    
In practice, however, the quantum fluctuations and finite amount of disorders in the real system push the actual solid-liquid line ($H_{\rm irr}$ and the solid line in the inset) down to lower magnetic fields, for example in the present material, down to about 4 T even at $T \simeq$ 0 K.  
Thus the $H_{\rm L}$ line is considered to have the similar nature of the remnant solid-liquid line at the original position where the short-range order of vortices starts to grow, but it does not develop to the long-range order with zero resistivity.
In addition the resulting vortex liquid is considered to be QVL at low temperatures.  
The low resistance state is, therefore, concluded to be a novel {\it quantum vortex slush} state. 

The $H_{\rm L}$ line seems to exceed the ordinary $H_{\rm c2}$ line at $T \simeq$ 0 K.  
This looks to be inconsistent with the above scenario of $H_{\rm c2} = H_{\rm L}$ at $T =$ 0 K. 
The Lindemann-type approach for the vortex melting theory leads the melting point always below the $H_{\rm c2}$ at $T =$ 0 K. \cite{Blatter94B}
Recently Ishida and Ikeda \cite{Ishida01} argue the melting transition near $T =$ 0 K by their quantum Ginzburg-Landau approach.  
They suggest that the melting transition is possible to occur above the ordinary $H_{\rm c2}$ at $T \simeq$ 0 K in the clean system and the quantum dissipated three dimensional case.  
It must be theoretically important to investigate the relation among $H_{\rm L}$, $H_{\rm irr}$, the ordinary $H_{\rm c2}$, and the mean field $H_{\rm c2}$ at $T \simeq$ 0 K.

\begin{figure}
\includegraphics[viewport=2cm 2cm 17cm 28.5cm,clip,width=0.9\linewidth]{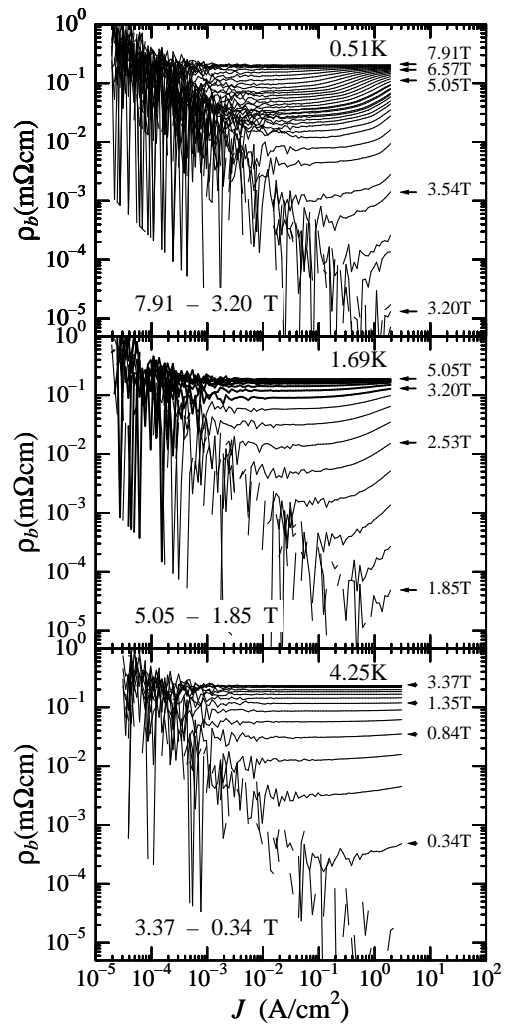}
\caption{Current density dependence of the in-plane resistivity $\rho_{b}$ of sample \#2 in magnetic fields perpendicular to the Q2D plane.  The curves are measured in intervals of about 0.17 T from 7.91 (top curve) to 3.20 T (bottom), from 5.05 to 1.85 T, and from 3.37 to 0.34 T at 0.51, 1.69, and 4.25 K, respectively.}
\end{figure}

Turning now to investigate the vortex slush like state characterized by the non-linear transport behavior.
Figure 5 shows the current density dependence of the in-plane resistivity $\rho_{b}$ of sample \#2 in magnetic fields perpendicular to the Q2D plane at 0.51, 1.69 and 4.25 K.  
The current density $J$ - resistivity $\rho_{b}$ characteristic curves are measured in intervals of about 0.17 T from 7.91 (top curve) to 3.20 T (bottom), from 5.05 to 1.85 T, and from 3.37 to 0.34 T at 0.51, 1.69, and 4.25 K, respectively.  
At the low temperature of 0.51 K (top panel), two regions of the constant $\rho_{b}$ as a function of $J$, that is, the linear $E$ - $J$ behavior, at low current density is separated at $H_{\rm L} \simeq$ 6.5 T. 
One is the liner resistivity in high magnetic fields near $H_{\rm c2}$, and another linear resistivity appears around 6 T.
In between these two regions, the steep drop of the linear resistivity at low $J$ is found as spreading the spacing between curves.  
Then below about 6 T, the linear resistivity stops dropping as fast. 
Only at lower magnetic fields approaching $H_{\rm irr}$, the downward curvature of the $\rho_{b}$ - $J$ curve sets in and the linear part of the resistivity vanishes within the present experimental accuracy.  
This indicates that the low resistance state below $H_{\rm L}$ has non-zero linear resistivity at low $J$.  
At high $J$, the low resistance state becomes unclear and shows the non-linear $E$ - $J$ behavior, which has been seen in Fig. 2.  
This non-linear behavior at high $J$ can be also explained in the vortex slush scheme.  
The vortex slush resistivity at low $J$ increases to the vortex liquid resistivity at high $J$ by moving the vortex lattice domains with short-range order due to the Magnus force.  
The domains pass over the pinning sites.  
For larger $J$, the vortex domains move like as vortex liquid, and hence a higher resistivity is induced.  
Actually as is seen in the top panel at 0.51 K, the smaller $J$ required for moving the domain, namely, where the $\rho_{b}$ - $J$ curve starts to deviate from the constant, is necessary with coming close to the magnetic field $H_{\rm L}$ (or $T_{\rm L}$).  
  At high temperature of 1.69 K (middle panel), where is just outside of the vortex slush like region, the linear resistivity at low $J$ shows continuous drop and then changes to the non-linear one with the downward curvature near $H_{\rm irr}$.  
Only the linear resistivity is observed within the applied $J$ in this experiments at 4.25 K (bottom panel).  

As just described above, the observed $E$ - $J$ response in the low resistance state is well explained by the concept of the vortex slush which has been proposed experimentally by Worthington {\it et al.} \cite{Worthington} and theoretically discussed by Ikeda, \cite{Ikeda01} although both studies have focused the attention on the oxide high-$T_{\rm c}$ superconductors. 
These studies have been examined also by the Monte Carlo simulations. \cite{Nonomura} 
While most of the characteristic feature is very well understood in the vortex slush concept, some points remain to be unclear.  
First, the linear resistivity in low $J$ in the vortex slush has been expected to come from thermal excitation, and then the exponential decrease of $\rho$ with $T$ has been predicted. \cite{Worthington} 
This exponential temperature dependence contradicts the observed weak temperature dependence as shown in Fig. 3.  
Second, the vortex slush state in the oxide high-$T_{\rm c}$ superconductors has been found only on the sample with the intermediate range of the disorders which have been controlled by the proton irradiation \cite{Worthington} and the oxygen concentration. \cite{Nishizaki00A,Shibata02} 
The organic superconductor in the present study is generally considered to be a clean system but with a finite weak disorder.  
These must closely connect to the relation among the origin (thermal or quantum) of the excitations (fluctuations), those strength at the temperature, and a degree of the finite pinning strength.  
Theoretical approach for the quantum effect on the vortices is necessary for further understanding of the present phenomena.  

Finally we would like to mention the $H_{\rm L}$ ($T_{\rm L}$) transition from the different aspects of the magnetic quantum oscillations.  
The dHvA oscillations have been observed not only in the normal state but also the vortex state below $H_{\rm c2}$. \cite{Wel,Sasaki98,Clayton}  
The amplitude of the oscillations in the vortex state has been reduced by an additional damping effect with respect to the normal state damping. \cite{Maniv}
The dHvA effect in the vortex liquid state of the present organic superconductor has been understood in the thermal fluctuation approach. \cite{Clayton,Maniv} 
Recently SdH oscillations in the vortex liquid state have been reported.  \cite{Sasaki02A}
The SdH oscillations have been able to be observed down to 5 T at 0.5 K, where the resistivity becomes about 30 \% of the normal state value.  
The additional damping of the SdH amplitude has appeared below about 7 T as well as the dHvA oscillations.  
In addition the novel second damping has been found only in the SdH effect below about 6 T at 0.5 K.  
The second damping suppresses the SdH oscillation amplitude rapidly as compared to the additional damping below $H_{\rm c2}$.
In view of the vortex phase diagram in the present study, the magnetic field where the second damping starts to appear corresponds to the $H_{\rm L}$ line.  
The amplitude of the dHvA oscillations in the vortex state has been calculated to be perturbed by the phase coherence of the vortices. \cite{Maniv} 
In the vortex slush state the coherence may vary with the applied current.  
The current dependence of the phase coherence in the vortex slush state may explain that the amplitude of SdH oscillations is smaller than that of the dHvA one below the second damping field $\simeq$ $H_{\rm L}$.
It must be, however, necessary for further consideration on the different way of the amplitude damping of both the dHvA and SdH effects in the vortex liquid (QVL or TVL) and the vortex slush states in addition to the normal and the Abrikosov vortex lattice states.

\section{Summary}

We have reported the non-linear low resistance state in the vortex liquid state of the quasi-two dimensional organic superconductor $\kappa$-(BEDT-TTF)$_{2}$Cu(NCS)$_{2}$.  
The low resistance state appears below about 1 K, which is separated from the thermal vortex liquid state by the characteristic drop of the in-plane resistivity at $T_{\rm L}$ and $H_{\rm L}$.  
A possible origin of the drop of the resistivity is a hidden freezing transition which is obscured by strong quantum fluctuations.

The low resistance state at lower temperature persists down to $T \simeq$ 0 K in $H_{\rm irr} < H < H_{\rm c2}$.  
The finite resistivity remained at $T \simeq$ 0 K and the reversible magnetization demonstrate that the quantum vortex liquid state is realized there.  
The drop of the resistivity at $T_{\rm L}$ and $H_{\rm L}$ comes from the linear resistivity at low $J$ in the low resistance state.  
By applying high $J$ the linear resistivity changes to the larger vortex liquid resistivity with non-linear response.  
These transport phenomena are well understood in the vortex slush concept, which is characterized by a short-range order of the vortices.  
As a result, the low resistance state is considered to be a novel {\it quantum vortex slush} state.
In order to confirm the quantum vortex slush in the low resistance state, it is important to investigate the thermodynamic properties at $H_{\rm L}$ and the way of the connection between the $H_{\rm L}$ and $H_{\rm irr}$ lines.

\begin{acknowledgments}

The authors thank R. Ikeda and T. Maniv for stimulating discussions. 
One of the authors (T. S.) acknowledges the support of a dilution refrigerator experiment at WMI by K. Neumaier and W. Hehn.   
A part of this work was performed at HFLSM, IMR, Tohoku University.  
This work was partly supported by a Grant-in-Aid for Scientific Research from the Ministry of Education, Science, Sports, and Culture of Japan.  

\end{acknowledgments}

\end{document}